\documentclass[two column,prb]{revtex4}
\usepackage{bm}
\usepackage{graphics}
\usepackage{graphicx}
\usepackage{amsfonts}
\usepackage{amsmath}
\usepackage{amssymb}
\usepackage{graphicx}
\usepackage{color}

\begin{document}

\preprint{HEP/123-qed}
\title[Short title for running header]{Enthalpy of formation for Cu-Zn-Sn-S (CZTS)}
\author{S.V. Baryshev$^{1,2}$}
\email{sergey.v.baryshev@gmail.com}
\author{E. Thimsen$^3$}
\email{elijah.thimsen@gmail.com} \affiliation{$^1$Euclid TechLabs
LLC; $^2$High Energy Physics Division, Argonne National Laboratory
\\ $^3$Department of Chemical Engineering and Materials Science,
University of Minnesota}

\begin{abstract}
Herein we report an analytical procedure to calculate the enthalpy
of formation for thin film multinary compounds from sputtering
rates measured during ion bombardment. The method is based on
Sigmund's sputtering theory and the Born-Haber cycle. Using this
procedure, an enthalpy of formation for a CZTS film of the
composition Cu$_{1.9}$Zn$_{1.5}$Sn$_{0.8}$S$_4$ was measured as
--930$\pm$98 kJ/mol. This value is much more negative than the sum
of the enthalpies of formation for the constituent binary
compounds, meaning the multinary formation reaction is predicted
to be exothermic.\\
(Supporting information is available at
\texttt{\textcolor[rgb]{0.00,0.00,1.00}{http://arxiv.org/abs/1403.4496}})
\end{abstract}

\maketitle

Cu$_2$ZnSnS$_4$ (CZTS) has generated tremendous interest as an
earth abundant, low-cost alternative to Cu(In,Ga)Se$_2$ (CIGS),
which is one of the key photoabsorber materials in commercial thin
film solar cells. As is often the case, the potentially low-cost
alternative presents challenges in terms of performance. For CZTS,
great emphasis was initially placed on getting the best device
performance, and power conversion efficiencies have rapidly
saturated at 9--11\%.\cite{1,2} The direction is now to improve
the material to reach the 20\% power conversion efficiency
mark.\cite{2} It is believed that to reach that high performance
level, CZTS must be better understood at a fundamental level. One
fundamental parameter that remains experimentally unknown is the
standard enthalpy of formation. There have been two values
reported that were calculated using density function theory (DFT);
--337 kJ/mol reported by Maeda et al.\cite{3} and --406 kJ/mol
reported by Walsh et al.\cite{4} However, to our knowledge, no
experimental measurements have been made. The small sample mass of
thin films makes measurement of thermochemical properties using
traditional techniques challenging. Thus there is a need for
alternative techniques that can measure small amount of sample.

Herein we report an approach to calculate the standard enthalpy of
formation from measured relative sputtering rates under ion
bombardment in the low energy regime (surface binding energy $\ll$
projectile energy $<$ 1 keV). The method relies upon the use of an
internal standard that has known composition and enthalpy of
formation. Using Sigmund's formula for sputtering yield in the low
energy regime, the ratio of the sputtering rate of the internal
standard to the sputtering rate of the unknown material, both
measured at the same current density and ion energy, are used
together with the measured composition of the unknown material to
calculate the surface binding energy ($U_0$) of atoms in the
unknown. The Born-Haber cycle is then used to convert the surface
binding energy to the enthalpy of formation, given the measured
composition of the unknown sample.

We believe that the method can be applied using data from common
surface science tools such as x-ray photoelectron and Auger
electron spectrometers (XPS and AES) and different types of
secondary ion mass spectrometers (SIMS) if the instrument is
equipped with an ion mill so composition profiles can be measured.
Such measurements should be within the capabilities of many
laboratories around the world, and we hope to see the
thermochemical tables start to fill with data on other interesting
multinary compounds.

The composition profiles of three different samples were used for
this study. All three samples were prepared by atomic layer
deposition (ALD) and the composition profiles were measured by
time of flight (TOF) SIMS. Experimental details can be found in
previous reports.\cite{5,6} The samples were 92 nm CZTS
(Cu$_{1.9}$Zn$_{1.5}$Sn$_{0.8}$S$_4$) coated by either 20 nm of
ALD ZnO, 22 nm of ALD ZnS or nothing.

The three SIMS depth profiles, which were used to measure relative
sputtering rates, are presented in Fig.1. Each profile was
measured at the same Ar$^+$ current density and projectile energy
($E_p$=250 eV). The experimentally-measured sputtering rate was
calculated by dividing the layer thickness by the time it took to
sputter through it:
\[
SR_{exp}=\frac{d}{t} \text{, (1)}
\]
where $d$ is the known layer thickness and $t$ is the time it took
to sputter through it during the measurement of the composition
profile.

\begin{figure}[t] \centering
\includegraphics[width=6.7cm]{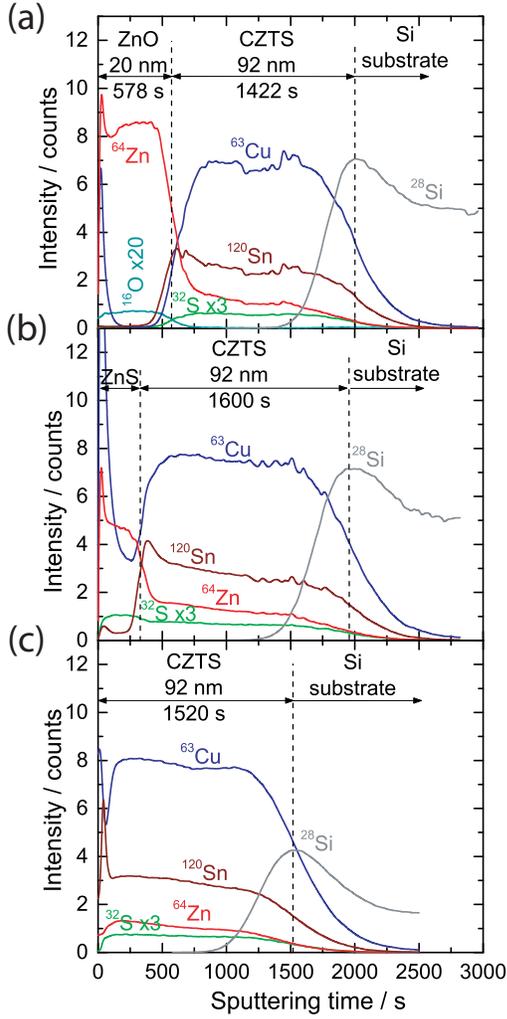}
\caption{\textbf{TOF SIMS composition profiles of 92 nm ALD CZTS
films coated with (a) 20 nm ALD ZnO; (b) 22 nm ALD ZnS; and (c)
uncoated.}}\label{fig:1}
\end{figure}

The sputtering rate, which has units of length per time, can be
written as:
\[
SR_{theory}=\frac{J\cdot Y}{e\cdot n} \text{, (2)}
\]
where $J$ is the ion current density, $Y$ is the sputtering yield
(i.e. the number of ejected atoms per incident ion), $e$ is the
elementary charge and $n$ is the atomic density of the target
material with units of atoms per volume. The atomic density can be
calculated if the composition and mass density are known by:
\[
n=\frac{\rho}{M_t} \text{, (3)}
\]
where $\rho$ is the mass density of the target material and $M_t$
is the number average atomic mass in the target material. $M_t$
can be calculated by:
\[
M_t=\frac{\sum \nu_i\cdot M_i}{\sum\nu_i} \text{, (4)}
\]
where $\nu_i$ is the number of atoms of element $i$ in the target
molecule and $M_i$ is the atomic mass of element $i$. The
sputtering yield can be calculated in the near-threshold (low)
energy regime using Sigmund's formula:\cite{7}
\[
Y=\frac{3}{4\pi^2}\cdot\frac{\alpha\cdot\gamma}{U_0}\cdot E_p
\text{, (5)}
\]
where $\alpha$ and $\gamma$ are functions of $M_t$ and the
projectile atomic mass, $E_p$ is the projectile energy and $U_0$
is the surface binding energy of atoms in the target material,
which has units of energy per atom. The surface binding energy
$U_0$ is the link to the thermodynamic properties of the target
material and that will be discussed later. The following formula
can be used to calculate $\gamma$:\cite{7}
\[
\gamma=\frac{4\cdot M_t\cdot M_p}{(M_t+M_p)^2} \text{, (6)}
\]
where $M_p$ is the atomic mass of the projectile, Ar$^+$ in this
case, and $M_t$ can be calculated using Eq.(4). Different authors
have reported different functions for the parameter $\alpha$, and
they agree to within 10\% of the value. We chose to use the
popular expression of Matsunami et al.:\cite{8}
\[
\alpha=0.08+0.164\cdot\Big(\frac{M_t}{M_p}\Big)^{0.4}+0.0145\cdot\Big(\frac{M_t}{M_p}\Big)^{1.29}
\text{. (7)}
\]

For two different materials bombarded by ions at the same
conditions, the ratio of the sputtering rate of material $2$ to
the sputtering rate of material $1$ can be predicted as a function
of the surface binding energy, atomic density and average atomic
mass by:
\[
\Big(\frac{SR_2}{SR_1}\Big)_{theory}=\frac{U_{0,1}}{U_{0,2}}\cdot\frac{n_1}{n_2}\cdot\frac{\alpha_2}{\alpha_1}\cdot\frac{\gamma_2}{\gamma_1}
\text{. (8)}
\]
If the composition and densities are known, then the only unknowns
on the right-hand-side in Eq.(8) are the surface binding energies
$U_{0,1}$ and $U_{0,2}$.

It is important to mention that it has been observed
experimentally that greater than 99.4\% of the gas-phase species
produced by Ar$^+$ bombardment of GaAs in the low energy regime
are neutral Ga and As atoms.\cite{9} Thus, sputtering of a
material, such as the inorganic sulfides and oxides considered
here, by Ar$^+$ in the low energy regime may be approximated as an
ideal process of generating gaseous atoms from the solid. The
energy penalty associated with promoting the atoms from the solid
into the gas phase is the surface binding energy, $U_0$.

The Born-Haber cycle is a theoretical thermodynamic cycle that
involves two different paths to transform atoms from their
condensed state in a solid into a liberated gas-phase state
(Fig.S5). One path involves first transforming the compound into
its constituent elements at their standard state, which requires
an energy change equal to the negative of the enthalpy of
formation. The second step is the generation of gaseous atoms from
the elements in their standard state, which involves vaporizing
condensed phases and breaking any bonds that may be present in
molecular elements (e.g. elemental sulfur or oxygen). The other
path from the solid to gaseous atoms treats the solid as a
collection of ions. The first step is to liberate the ions into
the gas phase from the crystal lattice, which involves an energy
penalty equal to the lattice energy. The second step is to
generate neutral gas-phase atoms from gas-phase ions, which
involves an energy change equal to the negative of the sum of the
ionization energies for the atoms in the compound. Putting the
preceding discussion into mathematical form, we may write:
\begin{equation}
\begin{split}
& E_{coh}= U_0\cdot\sum\nu_i= \\
& =-\Delta H_f^0+\Big(\sum\nu_i\cdot\Delta H_{vap,i}+ \\
& +\sum_{anions}\nu_i\cdot\eta_i\cdot
D^0_{molecule,i}\Big)=E_{lattice}-E_{ion} \text{, (9)}\nonumber
\end{split}
\end{equation}
where $E_{coh}$ is the cohesive energy of the molecule, which is
simply the surface binding energy of an atom in that material
multiplied by the number of atoms in a molecule, $\Delta H_f^0$ is
the enthalpy of formation, $\Delta H_{vap,i}$ is the enthalpy of
vaporization for element $i$, $\eta_i$ is the number of bonds per
atom in elemental molecules (i.e. $\eta$=1 for sulfur),
$D^0_{molecule,i}$ is the energy required to break a bond in an
elemental molecule (e.g. O$_2$ or S$_8$) in order to generate free
atoms, and $E_{ion}$ is the energy required to generate the ions
with the same formal charge as that in the material (e.g. Cu$^+$,
Zn$^{2+}$, Sn$^{4+}$, S$^{2-}$, O$^{2-}$, Mg$^{2+}$ and etc.).
$E_{ion}$ can be calculated using the following equation:
\begin{equation}
\begin{split}
& E_{ion}=\sum_{cations}\Big(\nu_{M_i}\cdot\sum_{j=1}^nI(M_i^{j+})\Big)+ \\
&
+\sum_{anions}\Big(\nu_{A_i}\cdot\sum_{j=1}^mI(A_i^{j-})\Big)\text{,
(10) \nonumber}
\end{split}
\end{equation}
where $\nu_{M_i}$ is the number of atoms of cation $i$ in a
molecule, $I(M_i^{j+})$ is the ionization energy of the process
$M_i^{(j-1)+}\rightarrow M_i^{j+}$, $n$ is the formal charge of
cation $i$ in the material, $\nu_{A_i}$ is the number of atoms of
anion $i$ in a molecule, $I(A_i^{j-})$ is the electron affinity of
the process $A_i^{(j-1)-}\rightarrow A_i^{j-}$ and $m$ is the
formal charge of anion $i$ in the material. An example calculation
can be found in the supporting information. Eq.(9) is the key
connection between the surface binding energy in the sputtering
process and tabulated thermodynamic data.

To verify that the model predictions agree reasonably well with
the experimentally-measured sputtering rates, Eqs.(8) and (9) were
used to predict the relative sputtering rates of materials with
known thermodynamic properties. This was performed by comparing
the sputtering rate of ZnO to MgO and also by comparing ZnS to ZnO
(see supporting information). The predicted sputtering rate of ZnO
is 2.1 times higher than MgO; while the experimentally measured
sputtering rate of ZnO is 2.0 times higher than MgO when measured
at the same conditions (supporting information), which is
excellent agreement considering there are no adjustable parameters
in Eqs.(8) and (9). Using the depth profiles in Fig.1, which were
all measured at the same ion bombardment conditions, the
sputtering rate of ZnS was measured to be 1.9 times higher than
ZnO. Using Eqs.(8) and (9) it is predicted that the sputtering
rate of ZnS would be 2.1 times higher than ZnO, again excellent
agreement considering there are no adjustable parameters in the
model. Thus we conclude that the model provides a fair
thermodynamic description of the sputtering process for these
ionic materials, and so next the model was used to measure the
unknown CZTS. For details see the supporting information.

The sputtering rate of CZTS (Cu$_{1.9}$Zn$_{1.5}$Sn$_{0.8}$S$_4$)
was measured to be 0.061 nm/s, while the sputtering rates of ZnS
and ZnO were measured to be 0.065 nm/s and 0.035 nm/s respectively
at the same conditions (Table 1). By rearranging Eq.(8) and using
the measured sputtering rate ratio, the surface binding energy of
CZTS was calculated to be 4.0 eV/atom by comparing to ZnO; and 3.6
eV/atom by comparing to ZnS. Averaging these values gives 3.8
eV/atom. Rearranging Eq.(9), using tabulated thermodynamic data
(see the supporting information), and $U_{0,CZTS}$ = 3.8$\pm$0.4
eV/atom, the enthalpy of formation for CZTS is --1.2$\pm$0.13
eV/atom, --9.6$\pm$1.0 eV/molecule or --930$\pm$98 kJ/mol of
Cu$_{1.9}$Zn$_{1.5}$Sn$_{0.8}$S$_4$ molecules.

\begin{center}
\textbf{Table 1. Summary of parameters used to calculate the
enthalpy of formation for CZTS.}
\begin{tabular}{c|c|c|c|c}
& Cu$_{1.9}$Zn$_{1.5}$Sn$_{0.8}$S$_4$ & ZnO & ZnS & Units \\
\hline
$SR$ & 0.061 & 0.035 & 0.065 & nm/s \\
\hline
$\rho$ & 4.6 & 5.6 & 4.1 & g/cm$^3$ \\
\hline
$d$ & 92 & 20 & 22 & nm \\
\hline
$M_t$ & 53.91 & 40.69 & 48.72 & a.m.u. \\
\hline
$n$ & 5.1$\times$10$^{22}$ & 8.3$\times$10$^{22}$ & 5.1$\times$10$^{22}$ & 1/cm$^3$ \\
\hline
$\alpha$ & 0.29 & 0.26 & 0.27 & -- \\
\hline
$\gamma$ & 0.98 & 1.0 & 0.99 & -- \\
\hline
$U_0$ & 3.8 & 4.2 & 3.1 & eV/atom \\
\hline
$E_{coh}$ & 31.2 & 8.4 & 6.2 & eV/molecule \\
\hline
$\Delta H^0_f$ & --1.2$\pm$0.13 & --1.80 & --1.07 & eV/atom \\
\hline
$\Delta H^0_f$ & --930$\pm$98 & --347 & --206 & kJ/mol \\
\end{tabular}
\end{center}

A conclusion that can be drawn from the measured enthalpy of
formation for multinary CZTS is that the reaction of the binary
metal sulfides to form the multinary compound is exothermic. The
reaction of the binary metal sulfides to form CZTS is
reversible,\cite{10} and can be written as:
\[
Cu_2S + ZnS + SnS_2\leftrightarrow Cu_2ZnSnS_4 \text{. (R1)}
\]
The sum of the enthalpies of formation for the binary compounds is
--437 kJ/mol (Table S4), which means the enthalpy change for
reaction R1 is approximately --490 kJ/mol. This contrasts with the
enthalpies of formation that have been reported using DFT
approaches. Maeda et al.\cite{3} and Walsh et al.\cite{4} reported
enthalpies of formation of --337 kJ/mol and --406 kJ/mol
respectively, both of which predict that reaction R1 is
endothermic with an enthalpy change of +100 kJ/mol and +31 kJ/mol
respectively. Differential scanning calorimetric (DSC)
measurements should be able to clearly determine whether reaction
R1 is exothermic or endothermic, but such measurements are outside
the scope of this communication.

Taking the analysis a step further, the Gibb's free energy of
reaction can be calculated using our experimental enthalpy and
those predicted by DFT. The free energy of reaction for R1 can be
written in terms of the energies of formation:
\[
\Delta G_r = \Delta G_{f,CZTS}-\Delta G_{f,binaries} \text{. (11)}
\]
The free energies of formation for the binary metal sulfides as a
function of temperature are known, and the expressions can be
found in the supporting information (Eq.(S10)). The free energy of
formation for CZTS as a function of temperature can be estimated
by:

\[
\Delta G_{f,CZTS}\approx \Delta H^0_{f,CZTS}-T\cdot\Delta
S^0_{f,CZTS} \text{. (12)}
\]
The variables in Eqs.(11) and (12) are known except for
$S^0_{f,CZTS}$ and $\Delta G_r$. Scragg et al. reported a free
energy of reaction for R1 $\Delta G_r$(823 K)=--22$\pm$6
kJ/mol.\cite{10} Thus we can solve Eqs.(11) and (12) for
$S^0_{f,CZTS}$ using $\Delta G_r$(823 K), and then plot the free
energy of reaction for R1 as a function of temperature (Fig.2).

\begin{figure}[t] \centering
\includegraphics[width=7.5cm]{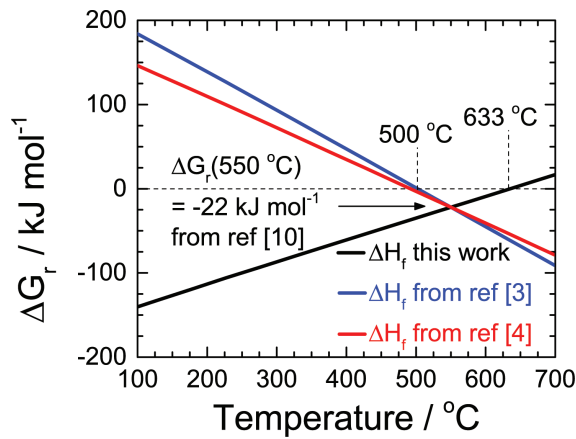}
\caption{\textbf{$\Delta G_r$ of reaction R1 using different
values of the enthalpy of formation from Eq.(11). The $\Delta
S^0_{f,CZTS}$ vales were --0.59, 0.13 and 0.046 kJ/(mol$\cdot$K)
using the enthalpies of this work, Ref\cite{3} and Ref\cite{4}
respectively.}}\label{fig:2}
\end{figure}

The enthalpy of formation measured in this work predicts a
different trend in the free energy of reaction when compared to
those predicted by DFT. An enthalpy of formation of --930 kJ/mol
predicts that the CZTS formation reaction from the binary metal
sulfides is favorable at low temperature, but becomes unfavorable
at approximately 633 $^\circ$C. Considering the margins of error,
the temperature at which we predict the reaction R1 to become
unfavorable is similar to the temperature of $\sim$550 $^\circ$C
where CZTS has been observed to decompose.\cite{10,11,12} The DFT
enthalpies of formation predict something quite different,
however, namely an endothermic reaction favorable at high
temperature since it is driven by an increase in entropy. Taken
with the free energy of formation reported by Scragg et al., the
DFT numbers predict the free energy of reaction to become negative
only at high temperature, with the line crossing the horizontal
axis at approximately 500 $^\circ$C (Fig.2). Experimentally,
multinary CZTS nanocrystals have been synthesized at temperatures
as low as 150--180 $^\circ$C.\cite{13} We have deposited a
Cu$_2$S/SnS$_2$/ZnS multialyer by ALD with an overall film
thickness of approximately 30 nm.\cite{14} After annealing for 60
minutes at 300 $^\circ$C in argon, this 30 nm multilayer structure
exhibited a Raman spectrum consistent with the multinary CZTS
phase, while it did not before annealing.\cite{14} Riha et al.
have synthesized CZTS nanocrystals at 300 $^\circ$C.\cite{15}
Summarizing, taken with the free energy of reaction reported by
Scragg et al.,\cite{10} the consequences of the enthalpy value we
have measured are 1) that the binaries should react to form the
multinary phase at low temperature (if kinetic limitations are
removed) and 2) that the multinary phase is unstable at 630
$^\circ$C (or even before because the reaction is reversible);
both of which have been experimentally observed.

There is a consequence for CZTS film processing. A number of
authors have reported on tin and sulfur loss from CZTS at elevated
temperatures that results from the SnS$_2$ (reaction R1)
decomposing into SnS and S$_2$, both of which are
volatile.\cite{10} The SnS$_2$ decomposition can be suppressed by
including sulfur or tin monosulfide in the annealing environment
at sufficiently high vapor pressure.\cite{10,12} Our hypothesis is
that a significant quantity of binary phases should be present at
high temperature, even if SnS$_2$ decomposition is suppressed.
However, since reaction R1 is reversible, and the multinary phase
is favored at low temperature (Fig.2), these binary phases should
react back into the multinary phase as the material cools, and
only the multinary phase should be observable at low temperature
for sufficiently slow cooling rates, provided SnS$_2$
decomposition has been suppressed. However, if the material is
rapidly quenched, we may expect that there will be significant
impurity content, since the binary compounds would have
insufficient time to react back into the multinary phase. In
short, for rapid quenching, binary impurities would be kinetically
frozen into the film.

Thanks to Melissa Johnson and Eray Aydil at UMN for useful
discussions about the content of the manuscript.

\end{document}